\documentclass{aastex}
\usepackage{spr-astr-addons}
\usepackage{url}

\RequirePackage{color}

\begin{document}

\title{Analytical model of Strange star in Durgapal spacetime}
\shorttitle{Short article title}
\shortauthors{Autors et al.}
\author{Rabiul Islam\altaffilmark{1}}
\affil{e-mail: rabiulphy19@gmail.com}
\author{Sajahan Molla\altaffilmark{2}}
\affil{e-mail: sajahan.phy@gmail.com}
\author{Mehedi Kalam\altaffilmark{3}}
\affil{e-mail: kalam@associates.iucaa.in}
\altaffiltext{1,2,3}{Department of Physics, Aliah University,
Action Area-II, New Town, Kolkata -700156, India.}

\begin{abstract}
A new strange star model based on Durgapal IV metric (Durgapal 1982) is presented here.Here we have applied a specific method to study the inner physical properties of the compact objects 4U 1702-429, 2A 1822-371, PSR J1756-2251,PSR J1802-2124 and PSR J1713+0747.The main objective of our study is to determine
central density ($\rho_{0}$), surface density ($\rho_{b}$), central pressure ($p_{0}$), surface redshift ($Z_{s}$), compactness and radius. Further we perform different tests to study the stability of our model and finally we are able to give an equation based on pressure and density i.e probable equation of state (EoS) which has an important significances in the field of Astrophysics.
\end{abstract}

\keywords{Stability; Mass-Radius relation; Compactness; Surface red-shift; Equation of stat; }


\section{Introduction}
Now a days people are very much interested to the study of dense objects(compact objects)in relativistic binary system.Neutron
Stars, Pulsars, Strange Stars are considered as compact objects.Neutron stars are made by mostly neutron particles , where as for strange stars, its strange quark particles. Actually in strange stars, up-down quarks are transformed to strange quarks which plays a crucial role at the core of the strange stars (Drago et al.2014;Haensel et al.1986).It is found that gravitational force of attraction is responsible for the stability of neutron star where as, strange stars are stabilized not only by gravitational force  but also by strong nuclear force. Therefore one can demand neutron star have lower force of attraction in  comparison  to strange star. Consequently, we may say that  the neutron star become larger in size over strange star  of same mass.It has been found that for neutron star surface matter density is almost zero, while for strange star it is not, at the boundary matter density will exist(Haensel et al. 1986; Alcock et al. 1986; Farhi and Jaffe 1986; Dey et al. 1998). After the formation of neutron star, its temperature has reduced to below the Fermi energy, hence the mass and radius of the neutron stars depends only on central density for a given equation of state. Also, it is very difficult to enumerate the mass and radius of that particular neutron star.For better conception one can go through the review work of Lattimer and Prakash(2007). It is to be mentioned here that for spherically symmetric static compact stars, theoretical calculation on mass and radius are the result of analytical
solution of Tolman-Oppenheimer-Volkov i.e., TOV equations.In order to study the astrophysical objects people used various method as  computational,
observational or theoretical analysis. Mass and radius of a compact star (neutron stars/strange stars) can be measured by pulsar timing, thermal emission from cooling stars, surface explosions
and gravity wave emissions, which are basically in observational point of view. Main objective is to obtain the proper equation of state in order to
describe the interior features of the neutron star(Lattimer and Prakash 2007; $\ddot{O}$zel 2006; $\ddot{O}$zel et al. 2009a; $\ddot{O}$zel and Psaitis 2009b;
G$\ddot{u}$ver and Cabrera-Lavers 2010a; G$\ddot{u}$ver et al. 2010b).
Although a very few number of compact star masses have been calculated accurately (to some extend) in binaries (Heap and  Corcoran 1992; Lattimer and Prakash 1992; Stickland and Lloyd 1997; Orosz and  Kuulkers 1999; Van Kerkwijk et al. 1995) but there is no information about the radius of that compact star.
Therefore,people has realized that  theoretical study is one of the important way for the study of stellar structure of newly observed masses and radius. Here, we want to mention some theoretical study on compact stars (Rahaman et al. 2012a; Rahaman et al. 2012b; Kalam et al. 2012; Kalam et al. 2013a;
Kalam et al. 2013b; Kalam et al. 2014a; Kalam et al. 2014b; Kalam et al. 2016; Kalam et al. 2017; Kalam et al. 2018; Hossein et al. 2012;
Jafry et al. 2017;Lobo 2006; Bronnikov and Fabris 2006; Maurya et al. 2016; Dayanandan et al. 2016; Ngubelanga and Maharaj 2015; Maharaj et al. 2014;
Paul et al. 2015; Pradhan and Pant 2014; Sharma et al. 2015).\\

N$\ddot{a}$ttil$\ddot{a}$ et al. (2017) has used the Rossi X-ray Timing Explorer Observations of five hard-state X-ray bursts to evaluate the mass of compact star in 4U 1702-429 and was found to be $1.9\pm 0.3 M_{\odot}$. Jhonker et al. (2003) gave another approach for the mass limit of neutron star  2A 1822-371 on the basis of phase resolved spectroscopic observations and pulse timing analysis and it was measured as $0.97 \pm0.24 M_{\odot} $.On the other hand, Ferdman et al. (2014) has found the mass of the pulsar PSR J1756-2251 as $1.341 \pm 0.007 M_{\odot}$. In another work, Ferdman et al. (2010) has measured (data from the
Parkes and Nancay observatories) the mass of the pulsar PSR J1802-2124 of the order of $1.24 \pm 0.11 M_{\odot}$. Splaver et al. (2005) has constrained the mass of the pulsar PSR J1713+0747 as $1.3\pm 0.2 M_{\odot}$ by using Shapiro delay. Here we are interested to find out the physical behaviour of the compact
objects like 4U 1702-429, 2A 1822-371, PSR J1756-2251, PSR J1802-2124 and PSR J1713+0747 by taking Durgapal (1982) spacetime metric, which described spherically symmetric and static configuration associated to an isotropic matter of fluid. \\

Outline of our study can be divided as follows: In section 2, we are dealing with interior space-time of the stellar compact objects. In section3 ,we have justified our model by keeping some test over Matter density and Pressure behavior, Energy conditions, Matching condition, TOV equation, Stability,
Adiabatic index,Mass-radius relation, Compactness, Surface red-shift in different subsection.  Finally overall discussion and concluding remarks with numerical data is introduced in section 4.
\section{Interior Spacetime of the compact object}
In spherically symmetric matter distribution of a compact object the interior spacetime as:
\begin{eqnarray}
ds^2 = -e^{\nu(r)}~dt^2 + e^{\lambda(r)}~dr^2 + r^2 (d\theta^2 +sin^2 \theta d\phi^2)
\end{eqnarray}
Here we consider that energy-momentum tensor for interior of the
compact object as,
\begin{equation}
               T_\nu^\mu=  ( -\rho , p, p, p)
\label{Eq2}
\end{equation}

where $\rho$ and $p$ are stands for energy-density and isotropic pressure respectively.\\
Accordingly, Einstein's field equations for the metric (1)  are
obtained as ($ G=1, c=1$)
\begin{eqnarray}
8\pi  \rho &=& e^{-\lambda}\left[\frac{\lambda^\prime}{r}-\frac{1}{r^2}\right]+\frac{1}{r^2},\label{eq2}\\
8\pi  p &=& e^{-\lambda}\left[\frac{\nu^\prime}{r}+\frac{1}{r^2}\right]-\frac{1}{r^2}.\label{eq3}
\end{eqnarray}
According to Durgapal(1982),
\begin{eqnarray}
e^{-\lambda} = \left(\frac{7-10Cr^2-C^2r^4}{7(1+Cr^2)^2}+\frac{KCr^2}{(1+Cr^2)^2(1+5Cr^2)^{2/5}}\right)\label{eq1}
\end{eqnarray}
\begin{eqnarray}
\nu =\ln \left(A(1+Cr^2)^4 \right) \label{eq1}
\end{eqnarray}
Where A(dimensionless), K(dimensionless) and \\
C($length^{-2}$) are constants. Solving the above equations, we get density($\rho$),
central density($\rho_0$), surface density($\rho_b$), pressure(p), central pressure($p_0$) as follows

\begin{eqnarray}
\rho = \frac{7CK(9C^2r^4-10Cr^2-3)}{56\pi(1+Cr^2)^3(1+5Cr^2)^{7/5}}  \nonumber \\ +\frac{8C(1+5Cr^2)^{2/5}(9+47Cr^2+11C^2r^4+5C^3r^6)}{56\pi(1+Cr^2)^3(1+5Cr^2)^{7/5}}
\end{eqnarray}
\begin{equation}
\rho_{0}=\frac{C(72-21K)}{56\pi}
\end{equation}
\begin{eqnarray}
\rho_{b}= \frac{7CK(9C^2b^4-10Cb^2-3)}{56\pi(1+Cb^2)^3(1+5Cb^2)^{7/5}} ~~~~~~~~~~~~~~~~~~~~~~~~ \nonumber \\ +\frac{8C(1+5Cb^2)^{2/5}(9+47Cb^2+11C^2b^4+5C^3b^6)}{56\pi(1+Cb^2)^3(1+5Cb^2)^{7/5}}
\end{eqnarray}
where $b = $ radius of the star.
\begin{eqnarray}
p=\frac{7Ck(1+9Cr^2)}{56\pi(1+Cr^2)^3(1+5Cr^2)^{2/5}} \nonumber \\
-\frac{16C(1+5Cr^2)^{2/5}(C^2r^4+7Cr^2-2)}{56\pi(1+Cr^2)^3(1+5Cr^2)^{2/5}} \nonumber \\
\end{eqnarray}
\begin{eqnarray}
p _{0}=\frac{C(32+7K)}{56\pi}
\end{eqnarray}
\section{Exploration of Physical properties}
In this section we will try to find out the following nature of the compact object:

\subsection{Density and Pressure Behavior of the compact object}
From Fig.~1 and Fig.~2 it is clear that, at the centre the density and pressure of the star is maximum and it decreases radially outward. Thus, the energy density and pressure are well behaved in the interior of the stellar structure. Interestingly, pressure drops to zero at the boundary, though density does not. Therefore, it may be justified to take these compact stars as a strange stars where the surface density remains finite rather than the neutron stars for which the surface density vanishes at the boundary (Haensel et al. 1986; Alcock et al. 1986; Farhi and Jaffe 1986; Dey et al. 1998). It is to be mentioned here that, we set the values of the constants $K(=0.00053)$ and $C(=0.00123 km^{-2})$, such that the pressure drops from its maximum value (at centre) to zero at the boundary.

\begin{figure}[htbp]
\centering
\includegraphics[scale=.3]{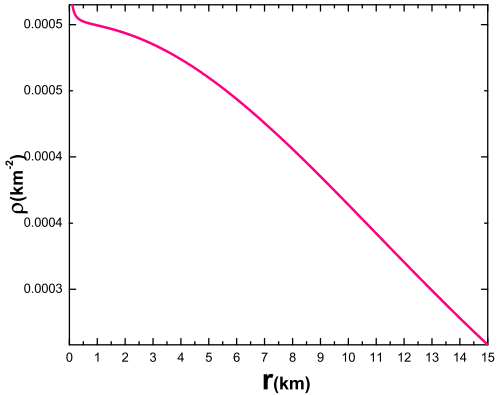}
\caption{Variation of the energy-density ($\rho$) at the stellar interior (taking K=0.00053 and C=0.00123$km^{-2}$).}
\label{fig:1}
\end{figure}
\begin{figure}[htbp]
\centering
\includegraphics[scale=.3]{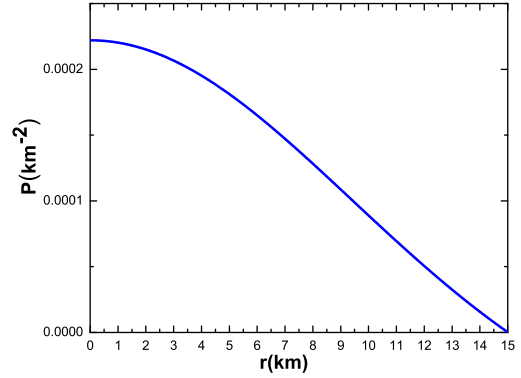}
\caption{Variation of the pressure ($p$) at the stellar interior (taking K=0.00053 and C=0.00123$km^{-2}$).}
\label{fig:2}
\end{figure}
\subsection{Energy conditions}
In our study, we have tested all the energy condition like null energy condition (NEC), weak energy condition (WEC), strong energy condition (SEC) and dominant energy condition (DEC) at the centre of the compact stars. Therefore from Fig.~1, Fig.~2 and Table~1 one can obtained the following energy conditions :\\
(i) NEC: $p_{0}+\rho_{0}\geq0$ ,\\
(ii) WEC: $p_{0}+\rho_{0}\geq0$  , $~~\rho_{0}\geq0$  ,\\
(iii) SEC: $p_{0}+\rho_{0}\geq0$  ,$~~~~3p_{0}+\rho_{0}\geq0$ ,\\
(iv) DEC: $\rho_{0} > |p_{0}| $.

\begin{table*}
\caption{Parameter values for the energy conditions (taking
K=0.00053 and C=0.00123$km^{-2}$)}
 \begin{tabular}{@{}cccc@{}}
\hline
 $\rho_{0}$ (km$^{-2}$) & $p_{0}$ (km$^{-2}$) & $\rho_{0}$+$p_{0}$ (km$^{-2}$)  & 3$p_{0}$+$\rho_{0}$ (km$^{-2}$)    \\
\hline
 0.000503307 & 0.000223752 & 0.000727059 & 0.001174563 \\
 \hline
\end{tabular}
\end{table*}

\subsection{Matching Conditions}
Interior metric of the strange star will be matched to the Schwarzschild exterior solution at the boundary i.e., at $r=b$(radius of the star).
\begin{eqnarray}
ds^2 = - \left(1-\frac{2M}{r}\right)dt^2 +  \left(1-\frac{2M}{r}\right)^{-1}dr^2 \nonumber\\
+ r^2 (d\theta^2 +sin^2\theta d\phi^2)
\end{eqnarray}
For the continuity of the metric functions $g_{tt}$, $g_{rr} $ and $\frac{\partial g_{tt}}{\partial r}$ at the boundary, we get
\begin{eqnarray}
 \left(\frac{7-10Cb^2-C^2b^4}{7(1+Cb^2)^2}+\frac{Kcb^2}{(1+Cb^2)^2(1+5Cb^2)^{2/5}}\right) = \nonumber   \\
   1-\frac{2M}{b}.\label{eq13}
\end{eqnarray}
\begin{equation}
 A(1+Cb^2)^4= \left(1-\frac{2M}{b}\right)  .\label{eq14}
\end{equation}
Now from the equation ~(13), we get the compactification factor as
\begin{equation}
u= \frac{ M(b)} {b}= Cb^2\left(\frac{24+8b^2C-\frac{7K}{(1+5b^2C)^{2/5}}}{14(1+b^2C)^2}\right) \label{15}
\end{equation}

\subsection{TOV equation}
The generalized TOV equation for fluid distribution takes the
form as
\begin{equation}
\frac{dp}{dr} +\frac{1}{2} \nu^\prime\left(\rho
 + p\right)
= 0.\label{eq18}
\end{equation}
TOV equation reveals that the strange star remain at equilibrium position by effective gravitational forces ($F_g$) and
effective hydrostatic ($F_h$) forces.
\begin{equation}
F_h+ F_g  = 0,\label{eq21}
\end{equation}
where,
\begin{eqnarray}
F_g &=& -\frac{1}{2} \nu^\prime\left(\rho+p\right)\label{eq22}\\
F_h &=& -\frac{dp}{dr} \label{eq23}
\end{eqnarray}
Now Fig.~3 shows the equilibrium of gravitational and hydrostatic
forces in the stellar structure.
\begin{figure}[htbp]
    \centering
        \includegraphics[scale=.3]{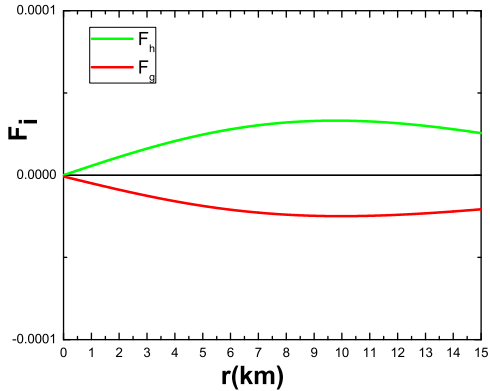}
        \caption{The gravitational ($F_{g}$) and hydrostatic ($F_{h}$) forces at the stellar interior (taking K=0.00053 and C=0.00123$km^{-2}$).}
    \label{fig:6}
\end{figure}


\subsection{Adiabatic Index}
In order to more tuning the model, one should check the infinitesimal radial adiabatic perturbation. Chandrasekhar (1964) had used this concept at first. Later this stability condition was used to various astrophysical cases by Bardeen et al. (1966), Knutsen (1988), Mak and Harko (2013). It is known fact that for
star modeling adiabatic index should be $\gamma = \frac{\rho+p}{p} \frac{dp}{d\rho}> \frac{4}{3}$.Therefore from Fig.~4, it is clear that $\gamma > \frac{4}{3}$ every point inside the strange stars.
\begin{figure}[htbp]

  \centering
        \includegraphics[scale=.3]{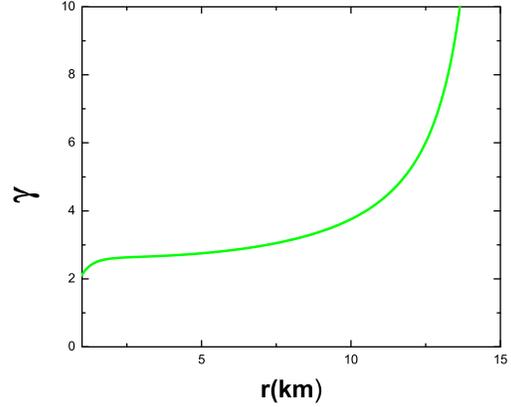}
       \caption{Variation of the adiabatic index $\gamma $ at the stellar interior (taking K=0.00053 and C=0.00123$km^{-2}$).}
   \label{fig:8}
\end{figure}
\subsection{Mass-Radius relation and Surface redshift}
In this subsection, we will try to investigate the maximum allowable mass-radius ratio of strange stars. Buchdahl (1959) has mentioned the mass-radius ratio limit for spherically perfect fluid sphere should be $\frac{ Mass}{Radius} < \frac{4}{9}$.In our study the relation between gravitational mass (M)  and energy
density ($\rho$) can be  written as
\begin{equation}
\label{eq34}
 M=4\pi\int^{b}_{0} \rho~~ r^2 dr =
Cb^3\left(\frac{24+8b^2C-\frac{7K}{(1+5b^2C)^{2/5}}}{14(1+b^2C)^2}\right)
\end{equation}
The compactness, u is given by
\begin{equation}
\label{eq35} u= \frac{ M(b)} {b}=
 C b^2\left(\frac{24+8b^2C-\frac{7K}{(1+5b^2C)^{2/5}}}{14(1+b^2C)^2}\right)
\end{equation}

Variation of mass function and compactness of the strange star are shown in Fig.~5 and Fig.~6.
\begin{figure}[htbp]
\centering
\includegraphics[scale=.3]{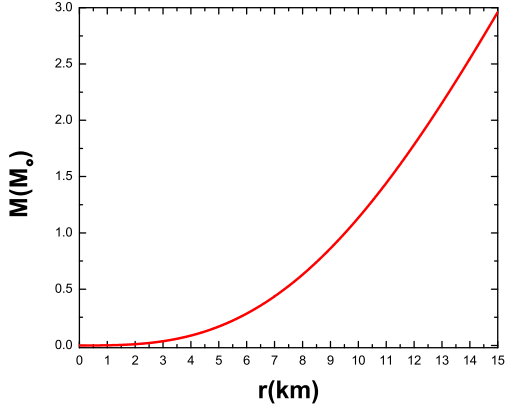}
\caption{Mass function at the stellar interior (taking K=0.00053 and C=0.00123$km^{-2}$).} \label{fig:9}
\end{figure}
\begin{figure}[htbp]
\centering
\includegraphics[scale=.3]{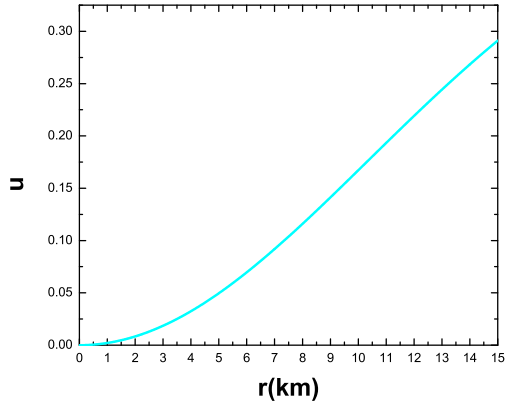}
\caption{Compactness (u) against radial parameter (r) at the
stellar interior (taking K=0.00053 and C=0.00123$km^{-2}$).}
\label{fig:10}
\end{figure}
\begin{figure}[htbp]
    \centering
        \includegraphics[scale=.3]{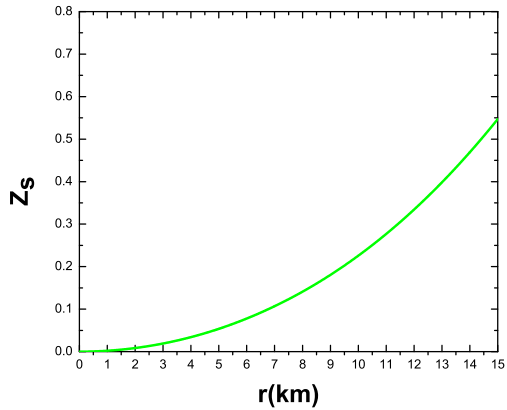}
        \caption{ Red-shift function ($Z_{s}$) against radial parameter (r) at the stellar interior (taking K=0.00053 and C=0.00123$km^{-2}$).}
    \label{fig:11}
\end{figure}

The surface red-shift ($Z_s$) due to the above compactness ($u$) is as follows:
\begin{equation}
\label{eq36} 1+Z_s= \frac{1}{\sqrt{1-\frac{Cb^2\left(24+8b^2C-\frac{7K}{(1+5b^2C)^{2/5}}\right)}{7(1+b^2C)^2}}}
\end{equation}
where
\begin{equation}
\label{eq37} Z_s= \frac{1}{\sqrt{1-\frac{Cb^2\left(24+8b^2C-\frac{7K}{(1+5b^2C)^{2/5}}\right)}{7(1+b^2C)^2}}} -1
\end{equation}
Thus, the maximum surface red-shift could be found easily from Fig.~7. The radii, compactness and surface red-shift of the
different strange stars are evaluated from Fig.~8, eqn.(21),eqn.(23) and a comparative analysis has been done in Table~2.
\begin{table*}
\caption{Radius, Compactness and Red-shift of strange stars are
mentioned below (taking K=0.00053 and C=0.00123$km^{-2}$).}
\scalebox{0.90}{
\begin{tabular}{@{}ccccc@{}}
\hline

 Star & Observed Mass($M_{\odot}$) & Radius from Model(in km) & Compactness from Model & Redshift from Model \\
 \hline
 4U 1702-429 & 1.9 $\pm$ 0.3  & 12.31 $\pm$ 0.84 & 0.240 $\pm$ 0.024 &0.394 $\pm$ 0.066\\
  2A 1822-371 & 0.97 $\pm$ 0.24 &9.33 $\pm$ 0.93 &0.155 $\pm$ 0.026 & 0.206 $\pm$ 0.045\\
 PSR J1756-2251 & 1.341 $\pm$ 0.007 & 10.67 $\pm$ 0.04 &0.193$\pm$ 0.001 & 0.276$\pm$ 0.002\\
  PSR J1802-2124 & 1.24 $\pm$ 0.11 & 10.35 $\pm$ 0.42 &0.184 $\pm$ 0.012 &  0.258 $\pm$ 0.024\\
  PSR J1713+0747 & 1.3 $\pm$ 0.2 & 10.51 $\pm$ 0.69 &  0.188$\pm$ 0.020  & 0.269$\pm$ 0.041\\
  \hline
\end{tabular}}
\end{table*}
\begin{figure}[htbp]
\centering
\includegraphics[scale=.2]{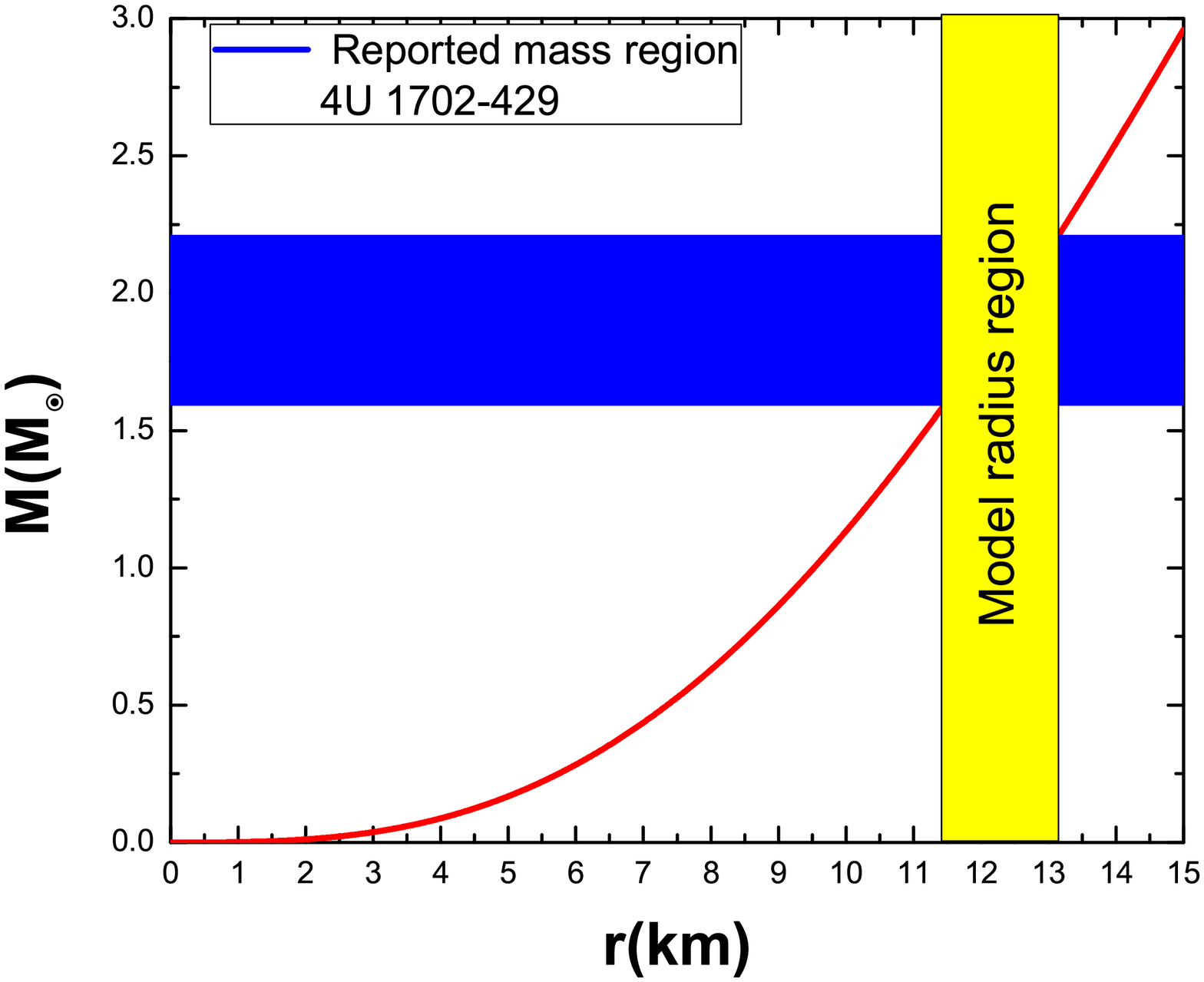}
\includegraphics[scale=.2]{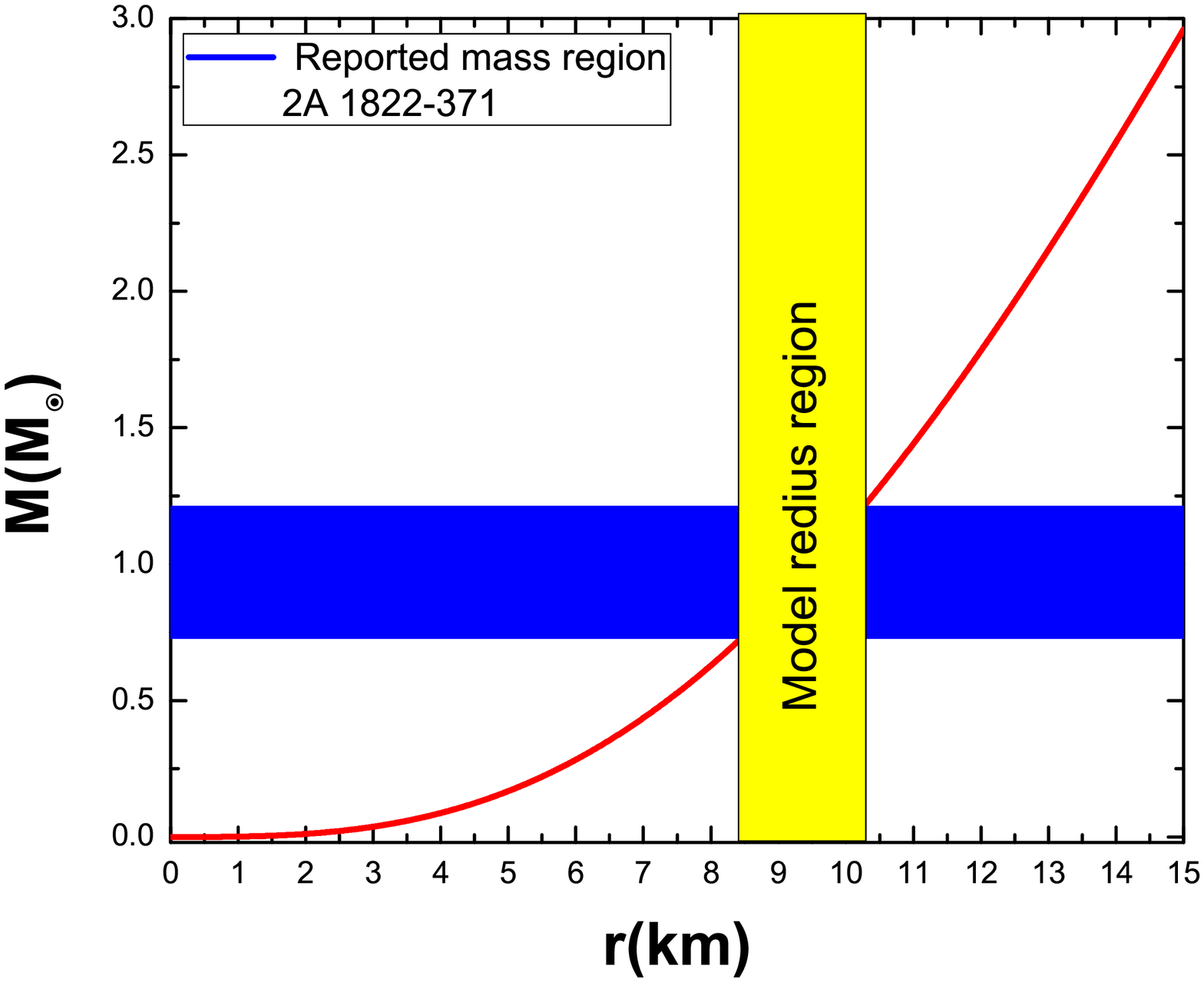}
\includegraphics[scale=.2]{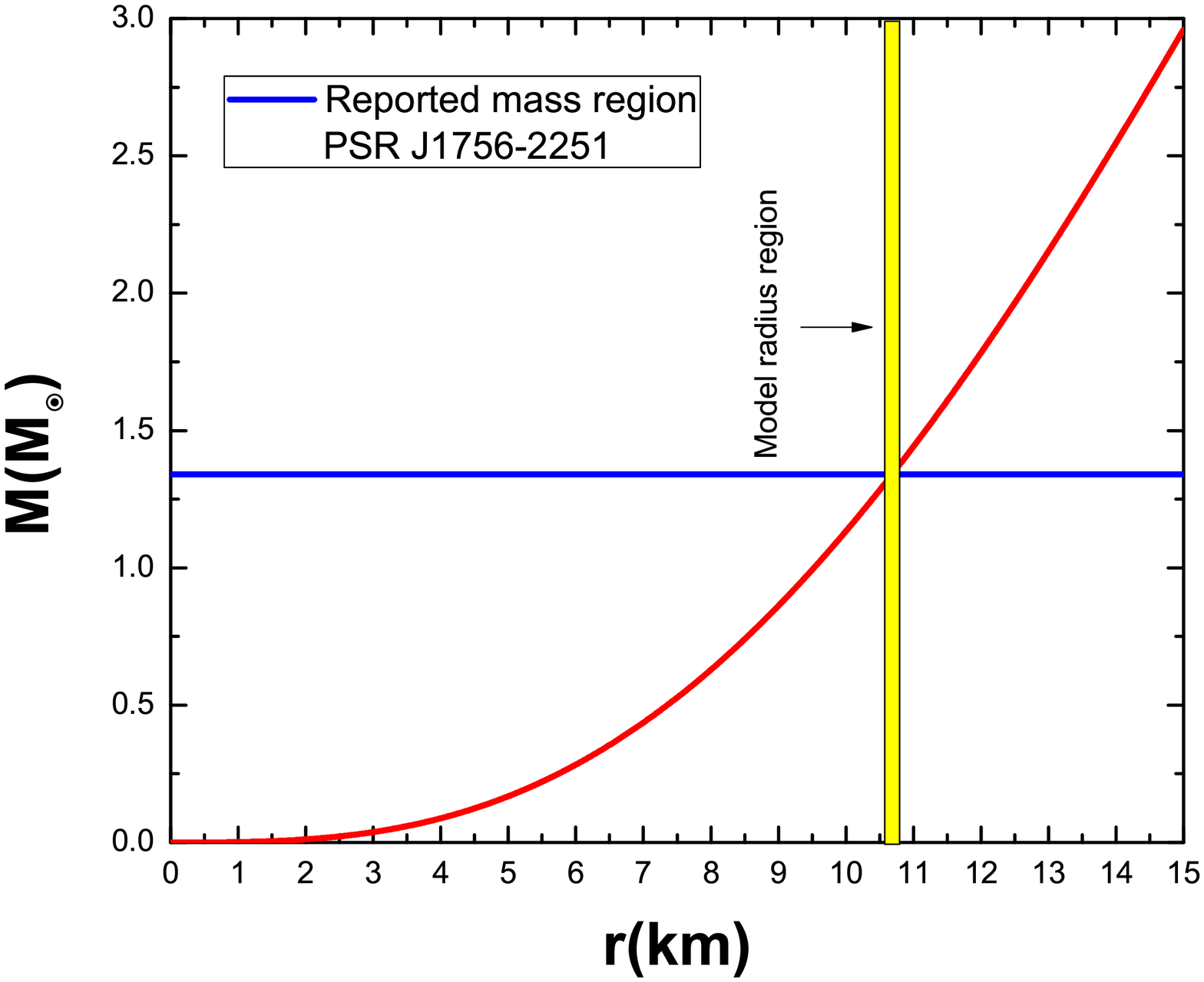}
\includegraphics[scale=.2]{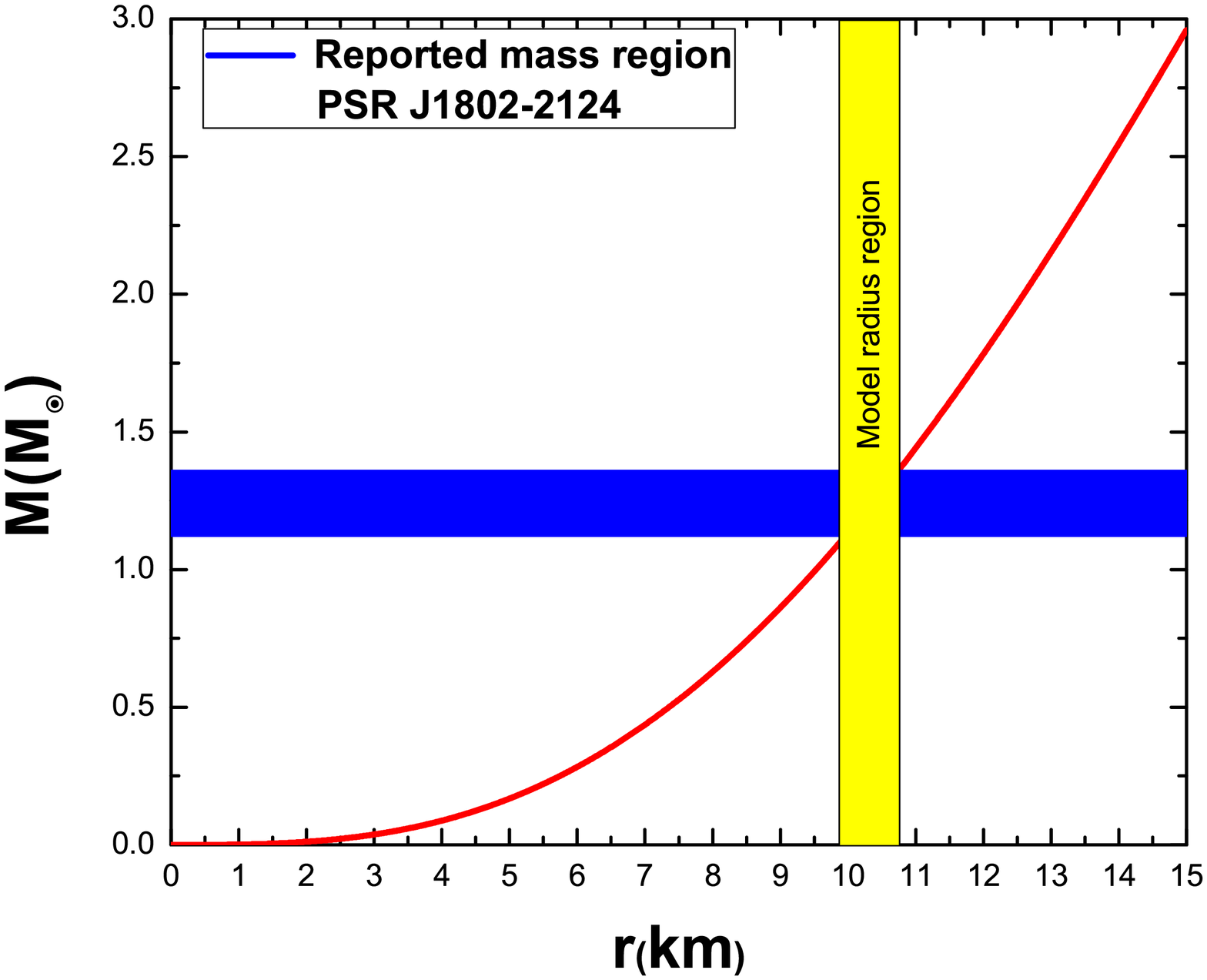}
\includegraphics[scale=.2]{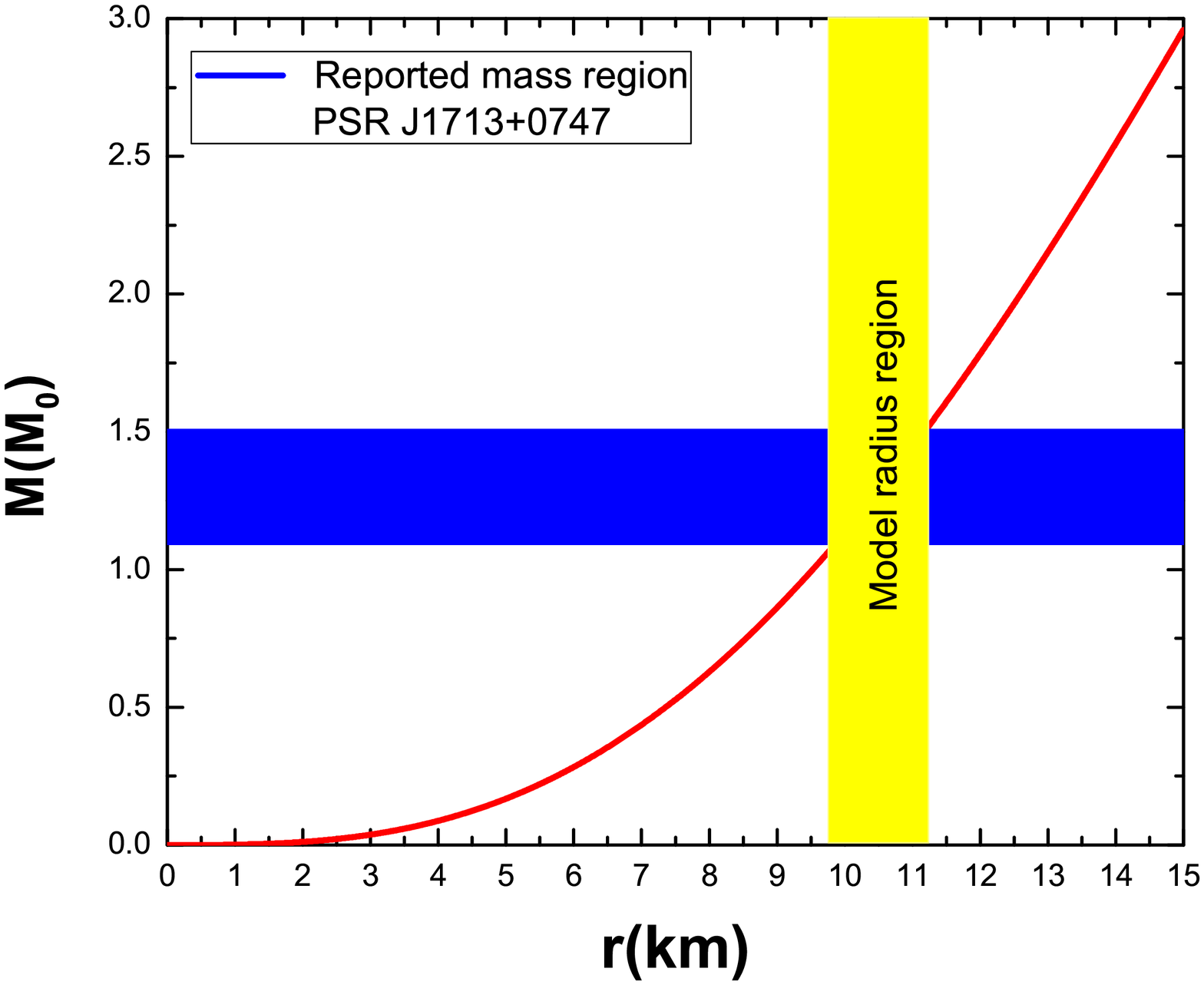}
\caption{Probable radii of 4U 1702-429, 2A 1822-371, PSR
J1756-2251, PSR J1802-2124 and PSR J1713+0747 (taking K=0.00053
and C=0.00123$km^{-2}$).} \label{fig:12}
\end{figure}
\section{Discussion and Concluding Remarks}
Based on present technique, we have proposed a new model of isotropic strange stars corresponding to the exterior Schwarzschild spacetime which is
singularity free. Here, we have studied several physical behaviour of the strange star namely 4U 1702-429, 2A 1822-371, PSR J1756-2251, PSR J1802-2124 and
PSR J1713+0747 under Durgapal (1982) IV metric spacetime.Conventionally, the mass-radius curve of compact stars are calculated under a given equation of
state for various values of central density; by a given value of the central density, the mass and radius of a compact star are fixed. It is to be mentioned
 here that our strange star model are different and theoretically interesting. According to our model, different strange stars are depends on the same
 parameter values (K=0.00053, C=0.00123$km^{-2}$) and consequently the same central density and the same equation of state. Therefore, interestingly,
if we starts from the centre with a certain central density, the model of a compact star can be determined by stopping at any radius where pressure
becomes zero. We think this model will give new dimension to study of compact stars.\\
Finally, we have got the variation of density and pressure at the interior of strange star in a systematic way. We observed that density and pressure are
maximum at the centre and gradually decreases as we move from centre to surface.Incorporating the value of G and c in the expression,we have calculated
the central
density($ \rho_0$) as $50.33~\times ~ 10^{-5}~ km^{-2}$ ($6.79~\times~ 10^{14}gm~cm^{-3}$) and
central pressure($p_{0}$) $22.37\times~10^{-5}~km^{-2} (5.58\times 10^{35}dyne/cm^2)$ (pl. see Table~1). We have verified all the energy conditions,
stellar equation (TOV) and stability conditions ( $\gamma = \frac{\rho+p}{p} \frac{dp}{d\rho}> \frac{4}{3}$). we have also matched our interior solution to
the exterior Schwarzschild line element at the boundary. From the mass function (eq.~20), the desired interior features of a strange star can be evaluated
which satisfies Buchdahl (1959) maximum mass-radius ratio. The surface redshift of the strange stars are found within standard value ($Z_{s}\leq 0.85$)
which is satisfactory (Haensel~2000). We have also obtained several physical parameters with numerical values as radius, compactness (u) and surface
redshift ($Z_s$) (pl. see Table~2) of the above mentioned strange stars. We have estimated the EoS and that is of the form $p=\alpha \rho + \beta$
where $\alpha(dimensionless)$ and $\beta(km^{-2}) $ are constants. According to our model, the estimated EoS (Fig.~9) should be a soft equation of state.
Therefore, we see that our analytical study of the isotropic strange stars: 4U 1702-429, 2A 1822-371, PSR J1756-2251, PSR J1802-2124 and PSR J1713+0747
under Durgapal (1982) IV metric satisfies all physical requirements of a stable strange star. As a consequence of this, we may conclude that one can find
useful relativistic model of strange stars under Durgapal IV metric by using the suitable choice of parameters K and C.
\begin{figure}[htbp]
    \centering
        \includegraphics[scale=.3]{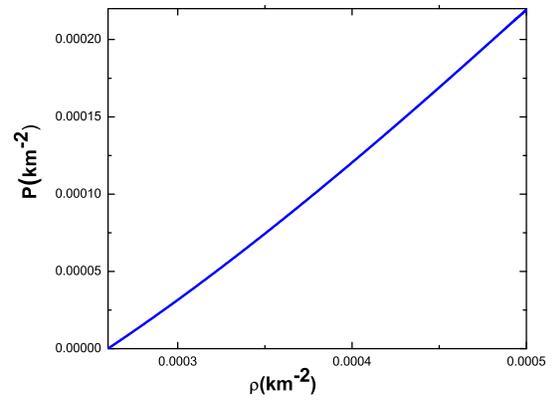}
        \caption{Possible pressure ($p$) - density ($\rho$) relation (EoS) at the stellar interior.}
    \label{fig:13}
\end{figure}

\acknowledgments MK would like to thank IUCAA, Pune, India and TWAS-UNESCO authority and IASBS, Iran for providing research facilities and warm hospitality under Associateship programme where a part of this work was carried out.RI would also like to thank UGC for giving UGC-MANF fellowship under which this research work has been done.

\end{document}